\documentclass[%
 reprint,
 amsmath,amssymb,
 aps,
prb
]{revtex4-1}

\usepackage{graphicx}
\usepackage{dcolumn}
\usepackage{bm}

\begin{document}


\title{Bimeron nanoconfined design}

\author{I. A. Iakovlev, O. M. Sotnikov, V. V. Mazurenko}
%
\affiliation{
Theoretical Physics and Applied Mathematics Department, Ural Federal University, Mira Street 19, Ekaterinburg 620002, Russia\\
}
\date{\today}

\begin{abstract}
We report on the stabilization of the topological bimeron excitations in confined geometries. The Monte Carlo simulations for a ferromagnet with a strong Dzyaloshinskii-Moriya interaction revealed the formation of a mixed skyrmion-bimeron phase. The vacancy grid created in the spin lattice drastically changes the picture of the topological excitations and allows one to choose between the formation of a pure bimeron and skyrmion lattice. We found that the rhombic plaquette provides a natural environment for stabilization of the bimeron excitations. Such a rhombic geometry can protect the topological state even in the absence of the magnetic field.
\end{abstract}

\pacs{Valid PACS appear here}
\maketitle


{\it Introduction.} \--- A specific shape of an object plays an important role in nature. One of the famous examples is the avian egg shape, which significant variation in degree of asymmetry and ellipticity was recently related to flight adaptation of birds\cite{eggs}. Another fascinating example is exoskeletons of viruses that are in the forms of the Platonic solids such as icosahedron\cite{wilczek}, which provides the best and fastest connection of the subunits.
There are also numerous examples when the particular shape choice revolutionized different fields of science or technology. In the first point-contact transistor invented by John Bardeen and Walter Brattain \cite{belllabs} a triangular form of the contact with one sliced tip was realized. It is important, that final shape of the system is formed via trial and error process during evolution or engineering independently on the origin of the system.

Here we address the problem of the system shape choice on the level of the topologically-protected magnetic excitations \cite{skyrmion1, Bogdanov} that attract considerable attention due to the potential technological applications in spintronics. The focus in this field of research gradually shifts from the study of the bulk crystals characterized by infinite skyrmion lattices \cite{skyrmion2,skyrmion3} to confined geometries \cite{confined1,confined2} with isolated topological excitations that can be controlled by means of electric and magnetic fields \cite{Nagaosa}. A practical realization of the skyrmion-based device with a strongly confined nanodisk geometry was recently reported in Ref.\onlinecite{expdisk}. The authors of the work have demostrated switching between different stable skyrmionic states in a 160-nm-diameter FeGe nanodisk. These experimental results stimulated the theoretical search for other confined geometries\cite{Pepper}.

The scanning tunneling microscopy (STM) allows one to manipulate individual atoms deposited on the surface and provides access to a completely another scale of the confined magnetic geometries with size of several nanometers. In this sense clusters or plaquettes of magnetic atoms constructed by means of the STM \cite{Loth} on a surface can be considered as elementary unit cell for stabilization of the topologically-protected excitations.   

\begin{figure}[t] 
\center 
\includegraphics[width=\columnwidth]{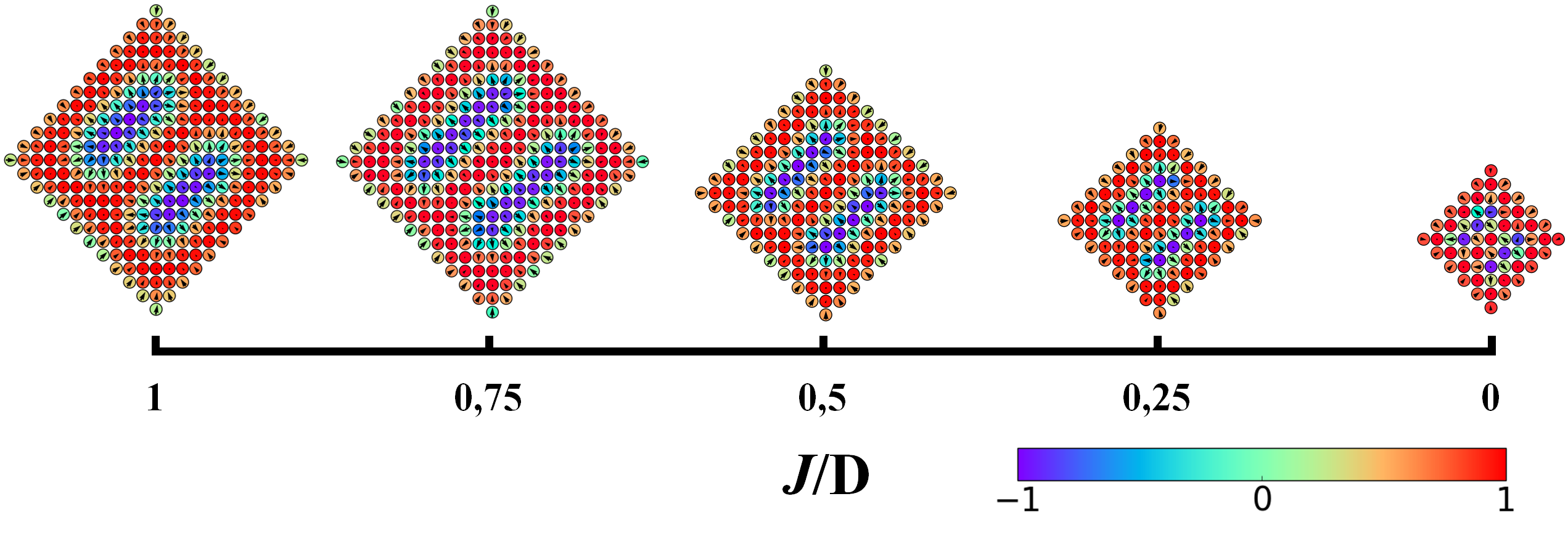} 
\caption{Family of the rhombic plaquettes with bimeron pairs stabilized in Monte Carlo simulations at different ratios $\frac{J}{D}$ ranging from 0 to 1. The magnetic field in these simulations was chosen to be 0.22, 0.3, 0.4, 0.7 and 1.3 for the plaquettes from left to right, respectively. The temperature is equal to 0.02. All the parameters are in units of DMI.}
\label{family} 
\end{figure}

In this paper we demonstrate that the choice of the rhombic shape of the spin plaquette is important to stabilize a distinct type of the topological excitations, bimerons that consist of two merons (half-disk domain carrying the skyrmion number $ Q= 1/2 $) separated by a stripe domain with zero topological charge. According to Fig.~\ref{family} the pair of bimerons can be stabilized on rhombic clusters in a wide range of the ratios between Dzyaloshinskii-Moirya interaction (DMI) and Heisenberg ferromagnetic exchange. Remarkably, being formed at the finite magnetic field these topological excitations remain stable with magnetic field switched off at very low temperatures, which is in demand for creating new atomic-memory technologies. At the same time, in the case of the two-dimensional square lattice the magnetic bimerons can be segregated in a fully controllable way with the vacancy grid. 

\begin{figure*}[t] 
\center 
\includegraphics[width=180mm ,clip]{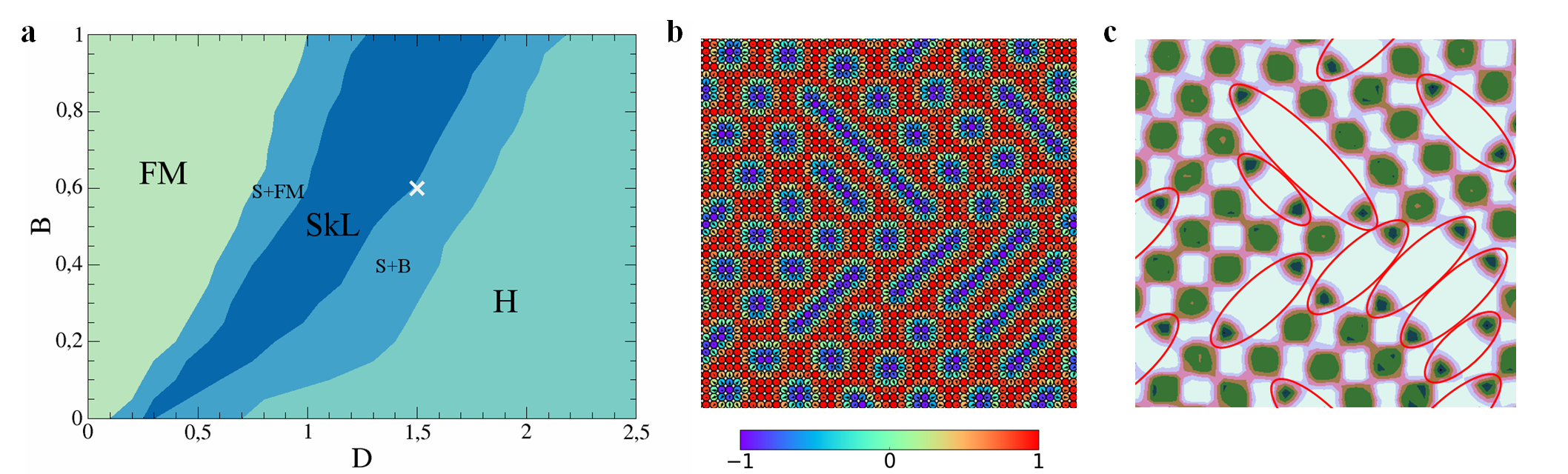} 
\caption{(a) Phase diagram of the ferromagnet with Dzyaloshinskii-Moriya interaction. The abbrevation SkL, S+FM, S+B and H denote skyrmion lattice state, non-periodic skyrmion state, mixed phase of skyrmions with bimerons and helical state respectively. The phase diagram was obtained at $T=0.02$. (b) Fragment of the square lattice with $L=512$ demonstrating the Monte Carlo solution obtained with the parameters $B=0.6$, $|{\bf D}|=1.5$, $T=0.02$ (white cross in (a)), which corresponds to the S+B phase. Magnetic atoms are represented by circles, black arrows denote the $xy$-plane projections of the spin moments, colour shows the $z$ component of the spin. (c) The calculated topological charge density. Bimerons are marked with red ovals. All the parameters are given in units of isotropic exchange interaction.}\label{phase}
\end{figure*}

{\it Model and Method.} \--- In our study we used the following spin Hamiltonian for simulations of the topological magnetic excitations on the L$\times$L square lattices:
\begin{equation}\label{Ham}
\begin{split}
H=- \sum_{i<j}J_{ij}{\bf S}_i{\bf S}_j-\sum_{i<j}{\bf D}_{ij}[{\bf S}_i\times{\bf S}_{j}]-  \sum_i B S_i^z
\end{split}
\end{equation}
where $J_{ij}$ and ${\bf D}_{ij}$ are the isotropic interaction and Dzyaloshinskii-Moriya vector, respectively. ${\bf S}_{i}$ is a unit vector along the direction of the $i$th spin and $B$ denotes the out-of-plane magnetic field. We take into account the interaction only between nearest neighbours. The isotropic exchange interaction is positive in our simulations, which corresponds to the ferromagnetic case. The symmetry of the Dzyaloshinskii-Moriya vectors is of C$_{4v}$ type, DMI has an in-plane orientation and perpendicular to the corresponding inter-site radius vector.

\begin{figure*} 
\center 
\includegraphics[width=180mm ,clip]{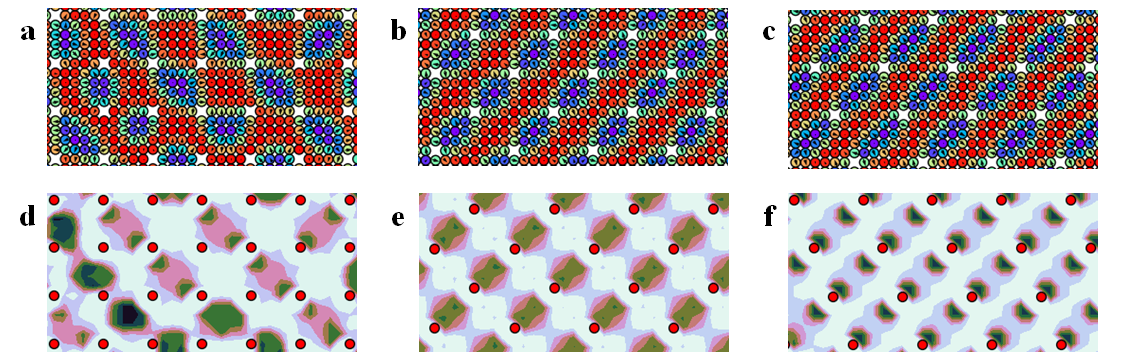} 
\caption{Fragments of the square lattice with $L=512$ demonstrating the Monte Carlo solution of the spin model, Eq.(\ref{Ham}) in presence of periodic vacancy grid with cells in the form of squares (a), rhombus (b) and parallelogram (c). (d)-(f) give the corresponding topological charge densities. Simulations were carried out at $B=0.4$, $J=0.67$, $T=0.02$ (white cross in Fig.~\ref{phase} (a)). All the parameters are given in units of DMI. }\label{regular_grid} 
\end{figure*}

To solve the spin Hamiltonian, Eq.(\ref{Ham}) we use the Monte Carlo approach with GPU parallelization \cite{cuda1,cuda2,cuda3}. It gives us opportunity to achieve significant acceleration of the Monte Carlo simulations up to 5-200 times compared to the CPU analogs ~\cite{cuda1,cuda2}. For example, we can perform simulations of large-scale two-dimensional systems with the linear sizes up to L=1024. Since each spin in our model interacts only with nearest neighbours, we chose the spin update scheme within the Metropolis algorithm is of checker-board type. During MC simulations, we gradually cooled down the system from high temperatures. The number of the temperature steps was equal to 50. Each run comprises 1.5 $\times 10^5$ MC steps per spin.

Since the values of the Dzyaloshinskii-Moriya interaction used in our simulations are equal or larger than the isotropic exchange interaction, then the resulting skyrmion species are compact. The possible realization of such an interaction regime was recently demonstrated in Ref.\onlinecite{Stepanov} by means of high-frequency laser fields. On the other hand there are surface nanosystems \cite{C2F} with $sp$ electrons that are naturally characterized by a strong supression of the isotropic interaction. To calculate the skyrmion number (topological charge) we adopted the approach proposed in Ref.\onlinecite{Berg}.

To identify the different phases realized in the system we used the calculated spin structure factor, topological charge denoted as Q and visualized a number of magnetic configurations during Monte Carlo simulations.

{\it Regular lattice.} \--- The phase diagram plot (Fig.~\ref{phase}\,(a)) shows that the skyrmion phase is composed of non-periodic skyrmion (S+FM) state, periodic skyrmion lattice (SkL) and mixed state of skyrmions and bimerons (S+B). Our main focus is on the S+B phase,  since we search for the conditions of pure bimeron excitations stabilization. A similar phase was simulated in Ref.~\onlinecite{Ezawa} with the two-dimensional non-linear sigma model and was experimentally observed at low temperatures in itinerant ferromagnets with Dzyaloshinskii-Moriya interaction \cite{natureYu}, however in the last work it was identified as a combination of skyrmions and fragments of helices. 

The width of observing bimerons corresponds to the diameter of the skyrmion. The latter is controlled by the $\frac{J}{|{\bf D}|}$ ratio. At the same time the mean length of bimerons increases with increase of DMI strength at a fixed magnetic field.

A typical example of the magnetic configuration corresponding to the S+B phase is presented in Fig.~\ref{phase}\,(b). The bimerons of different sizes are located along the diagonals of the square lattice with periodic boundary conditions. From the calculated skyrmion density plot, Fig.~\ref{phase}\,(c) one can see that each bimeron contains two merons and a rectangular stripe domain in the middle part. In such a system setup it is impossible to predict an exact length and location of a bimeron, which is our main interest in this study.  

\begin{figure}[!b] 
\center 
\includegraphics[width=\columnwidth ,clip]{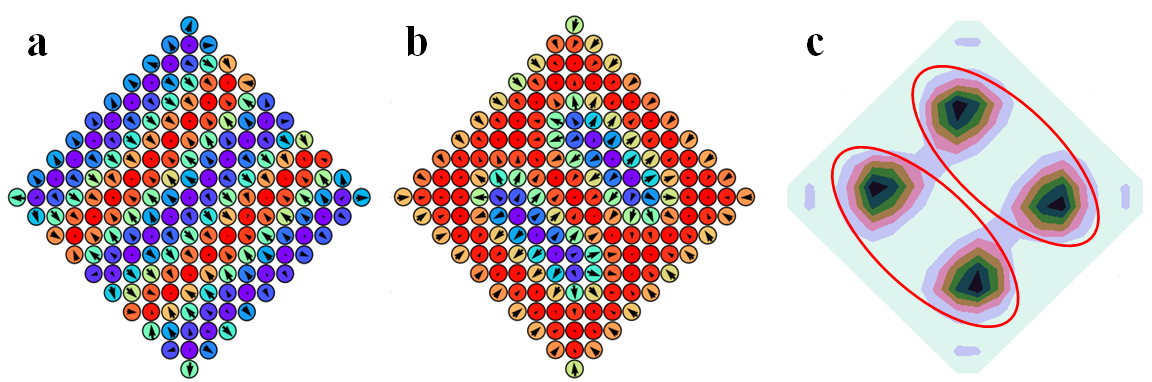} 
\caption{Monte Carlo solutions of the spin model, Eq.(\ref{Ham}) on the rhombic plaquette at $B=0$ (a) and $B=0.4$ (b). (c)~The topological charge density corresponds to (b). Bimerons are marked with red ovals. Here $J=0.67$, $T=0.02$. All the parameters are given in units of DMI.} \label{rhombicbimeron}
\end{figure}

\begin{figure}[b] 
\center 
\includegraphics[width=80mm ,clip]{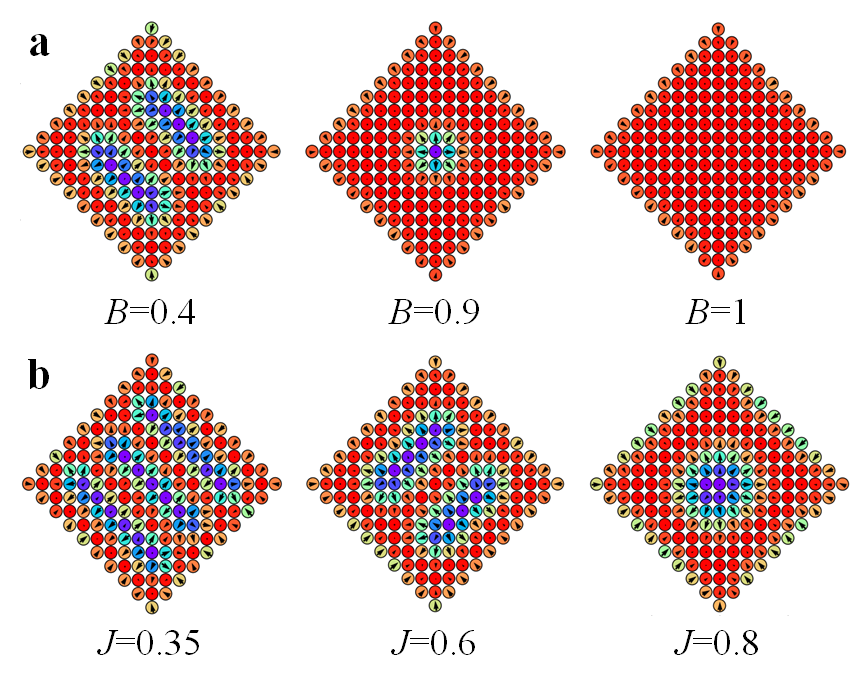}
\caption{(a) Dependence of the Monte Carlo solution on the magnitude of the magnetic field. Here $J=0.67$, $T=0.02$. (b) Dependence of the topological excitations on the $\frac{J}{D}$ ratio. Here $B=0.45$, $T=0.02$. All the parameters are given in units of DMI.}
\label{depend}
\end{figure}

{\it Vacancy grid.} \--- The effect of point defects on skyrmionic lattice state attracts considerable attention mainly due to the possibility to observe magnetic monopoles \cite{Milde} predicted by Dirac. Previous numerical studies\cite{Silva} on the discrete lattices revealed the formation of the bimerons in the presence of the vacancies randomly distributed over the system. The authors of Ref.\onlinecite{Silva} identified the mechanism of the bimeron stabilization as impact of effective local magnetic fields induced by vacancies. However, a probablistic formation of bimerons prevents one to use such a system in practical applications. In this work we mainly focus on the study of the topological excitations in systems where vacancies have regular positions, which opens a way for creation of atomic-scale memory units. In this respect, recent STM experiments \cite{kilobytememory} demonstrated an unprecedented possibility to manipulate the vacancies in the surface nanosystems for high-density storage of information. 

Here we used a similar idea to stabilize the bimerons on the two-dimensional lattice. To demonstrate the effect of vacancies insertion we choose the Hamiltonian parameters that correspond to the border between SkL and S+B phases. The corresponding Monte Carlo solution for the lattice without vacancies presented in Fig.~\ref{phase}\,(b) and Fig.~\ref{phase}\,(c) shows that the system is mainly filled by the skyrmions with several bimeron excitations. For the same set of parameters the regular square grid of vacancies created in the spin lattice leads to the formation of the bimerons of the same size. However, there is still a fraction of the skyrmions, Fig.~\ref{regular_grid}\,(a) and Fig.~\ref{regular_grid}\,(d). We found that the size of the square vacancy grid cell should be commensurate with the period of the spin spiral in the initial lattice without defects. For instance, in the case we consider the period of the spin spiral and the diagonal of the vacancies grid cell are equal to $4\sqrt{2}$ in the units of the lattice constant. With further increase in length, the bimerons also arise, but these configurations do not exhibit stable appearance of such excitations. 

A pure skyrmion or pure bimeron lattices can be stabilized by forming a vacancy grid with rhombic or parallelogram unit cell. As it is demonstrated in Fig.~\ref{regular_grid}\,(b) and Fig.~\ref{regular_grid}\,(c) the choice between pure bimeron and skyrmion phases can be made by choosing the corresponding shape of the vacancy grid cell. Importantly, the number of the magnetic atoms on the longer diagonal of the parallelogram is four, which corresponds to the period of spin spiral formed on the non-defective initial lattice at zero magnetic field.  

\begin{figure}[t] 
\center 
\includegraphics[width=80mm,clip]{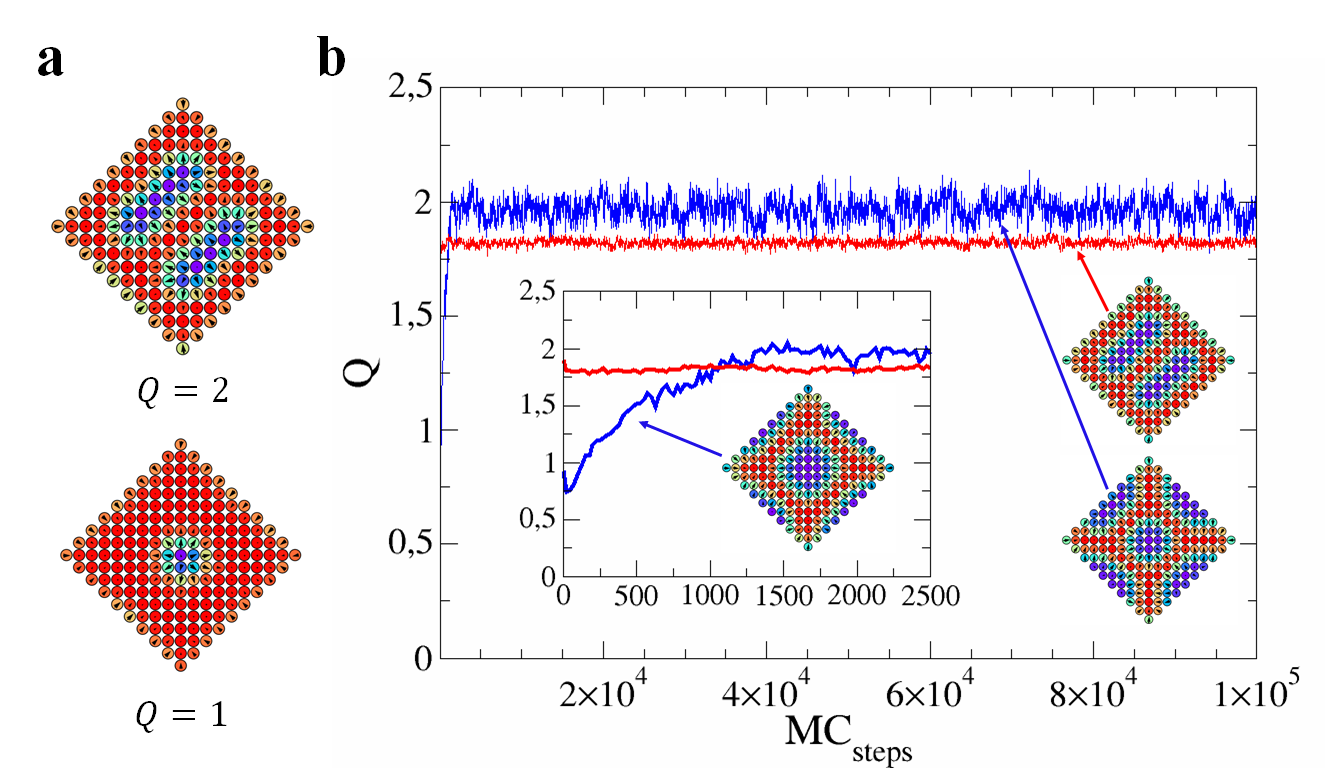} 
\caption{Demonstration of solution stability. (a) The bimeron (top) and skyrmion (bottom) magnetic configurations obtained at $B=0.4$ and $B=0.9$, respectively. Here, $T= 0.005$ and $ J = 0.67 $. (b) Time evolution of the prepared bimeron and skyrmion states after switching the field off. The inset shows short-time evolution of the topological excitations. All the parameters are given in units of DMI.} \label{stability}
\end{figure}

\begin{figure*}
\center 
\includegraphics[width=\textwidth]{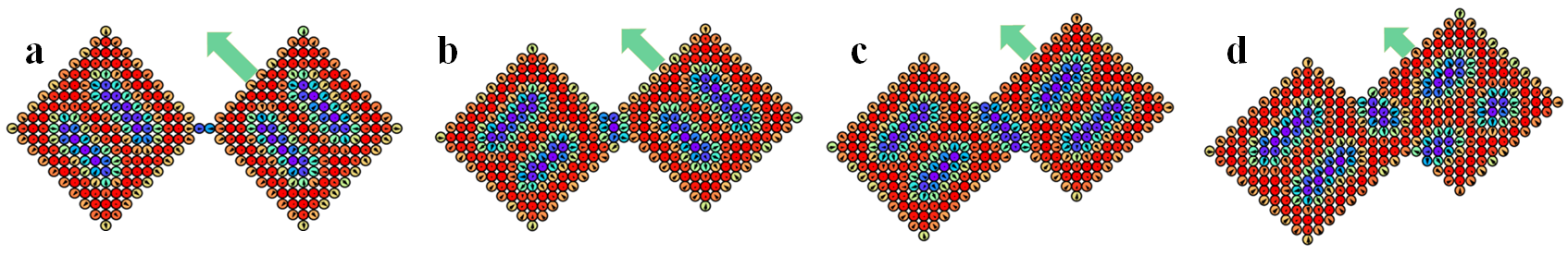} 
\caption{Example of connections between two rhombic plaquettes carrying pair of bimerons. The number of the contacting spins is 1, 3, 5 and 6  for (a) - (d) respectively. The parameters used in these simulations are $J=0.67$, $T=0.02$. The arrow denotes the direction of the shift of the plaquettes with respect to each other. All the parameters are given in units of DMI.} \label{connection}
\end{figure*}

{\it Rhombic plaquette} \--- Having simulated the lattices with periodic boundary conditions we are in a position to find a shape of a finite spin cluster in which the pure bimeron excitations are stabilized at finite temperatures and magnetic fields. Fig.~\ref{rhombicbimeron} demonstrates such a plaquette that is a rhombic fragment of the square lattice. At zero magnetic field we observe a spin spiral state, Fig.~\ref{rhombicbimeron}\,(a).  At constant ratio $J/D$, the plaquette state can be switched between bimeron pair, single skyrmion and fully polarized case by variation of the external magnetic field, Fig.~\ref{depend}\,(a). 

In turn, the different values of the isotropic exchange interaction at the fixed magnetic field of $0.45|{\bf D}|$ can produce three-bimeron, two-bimeron or skyrmion state, as it can be seen from Fig.~\ref{depend}\,(b).

One of the fascinating properties of the spin clusters with rhombic shape is that bimeron excitations remain stable when the magnetic field is instantly switched off. According to the Monte Carlo results presented in Fig.~\ref{stability}, the skyrmion number of the plaquette with starting bimeron configuration fluctuates around 1.8, which slightly smaller than the saturated value of 2 at the finite magnetic field. In turn, the single skyrmion stabilized at the magnetic field of 0.9$|{\bf D}|$ transforms into another structure with non-zero topological charge when the magnetic fields is switched off. As can be seen from Fig.~\ref{stability}\,(b), new structure is characterized by one skyrmion in the center of the plaquette and boundary spins that are antiparallel to the magnetic field. According to our simulation results, strong fluctuations of the topological charge are mainly related to fluctuations of the ortientation of the boundary spins. 

Since without magnetic field such a bimeron state corresponds to the local minimum of the system, it can be destroyed by temperature fluctuations. For instance, the systems visualised in Fig.~\ref{stability}\,(a) relax to a spin spiral state when the temperature is increased from 0.005 to 0.012$|{\bf D}|$ (for the bimeron pair state) or to 0.025$|{\bf D}|$ (for the skyrmion state). These results are sensitive to the details of the Monte Carlo simulations. For instance, in our scheme the new direction of a spin is choosen by using the solid angle restriction of 10$^{\circ}$. 

At last, the stability of the bimerons excitations in the connected rhombic clusters should be investigated in order to use such systems in real applications, for instance as a building blocks of nano-scale memory devices. The results for the two-plaquette configurations with different number of the boundary spins having neareast neighbours belonging to another plaquette are visualized in Fig.~\ref{connection}. The stable bimeron configurations in both plaquettes exist while the number of the contacting spins is less than 6 for plaquettes having 19 diagonal spins. We have checked that the complete linking of two plaquettes leads to the formation of a mixed skyrmion-bimeron state.

{\it Conclusions.} \--- We have shown that the two-dimensional bimeron lattice can be stabilized in controllable way by means of the regular vacancies grid created in the spin lattice with competing Dzyaloshinskii-Moriya and isotropic exchange interactions. The size and shape of the vacancy grid cell define the type and length of the topological magnetic excitations in the spin system. In the limiting case of the finite spin cluster one needs to choose the rhombic shape to guarantee the formation of bimerons. The obtained results can be used to guide future scanning tunneling microscopy experiments aiming to control the topological excitations in confined nanostructures.

{\it Acknowledgements} \--- We thank Alexander Tsirlin for fruitful discussions.

\end{document}